\documentclass[a4paper,11pt]{article}
\usepackage{pos}

\title{Recent Hadron Spectroscopy Results from the Belle and Belle II experiments}

\author*[a]{Elisabetta Prencipe}

\affiliation[a]{Justus-Liebig University of Giessen,\\
  II. Physics Institute, \\
  Heinrich-Buff-Ring 16, 35392 Gießen, Germany}

\emailAdd{elisabetta.prencipe@gmail.com}

\abstract{
The Belle~II experiment, the major upgrade of the Belle detector, is currently collecting data at the SuperKEKB asymmetric-energy $e^+e^-$ collider in Tsukuba, Japan. With an integrated luminosity of 575\,fb$^{-1}$, Belle~II has already surpassed the maximum luminosity accumulated by both BaBar and Belle. 
Recent analyses combining Belle and Belle~II data sets have delivered results of unprecedented precision, demonstrating the strong potential of the unified data sample, which now exceeds 1.5\,ab$^{-1}$.

In this contribution, recent advances in hadron spectroscopy are presented. 
Highlights include new results in bottomonium and charmonium physics, with particular emphasis on the structure of the $\Upsilon(10753)$, searches for hidden-charm and hidden-strangeness pentaquarks, and the latest developments in baryon spectroscopy, including studies of the $\Omega(2012)$ and excited $\Lambda_c$ states. 
Prospects for future spectroscopy analyses exploiting the full Belle and Belle~II data sets are also discussed.
}

\FullConference{Proceedings of the 53$^{th}$ International Symposium of Multiparticle Dynamics at the Corfu Summer Institute 2025 "School and Workshops on Elementary Particle Physics and Gravity" (CORFU2025)\\
27 April - 28 September, 2025\\
Corfu, Greece\\}

\begin{document}
\maketitle

\section{Introduction}
Belle II, which is a major upgrade of the Belle experiment, is located in Tsukuba (Japan). It collected 575 fb$^{-1}$ integrated luminosity data, which in combination with the Belle data sets, form the biggest data sets ever collected at an antisymmetric electron-positron collider: a total data set of 1.5 ab$^{-1}$ integrated luminosity is available to perform studies of CP violation on B decays, measure rare branching fractions, studies in the dark sector, and charm, charmonium and bottomonium spectroscopy.
Multiquark states search is an area of great interest in the physics community. The Belle II experiments offers the possibility to search for multiquark states through different production mechanisms:
\begin{itemize}
\item direct production in $e^+ e^-$ collisions
\item production through B decays
\item photon-photon scattering
\item double charmonium.
\end{itemize}

The Belle experiment (1999-2010) has already given a remarkable contribution in spectroscopy search. Starting with the X(3872) \cite{ref_X3872}, and followed by the obervation at Belle of the Z(4430)~\cite{ref_Z4430_PRL2008, ref_Z4430_PRD2014}, they represent only two example over tens of four-quark states unluckily explained with the quark model, showing the high contribution given by the Belle collaboration in the field of charmonium(-like) spectroscopy. The advantage at Belle, as for all $e^+e^-$ colliders, is that events are basically background clean compared to $p \bar p$ detectors, a huge data set is available, and several mechanisms are available to perform spectroscopy studies.

We observe that the higher the mass, the more rare the process is. Therefore  huge data sets are needed to engage in specific spectroscopy campaigns. This motivated the need to bind the Belle and Belle II data sets, so we have so far more than 1.5 ab$^{-1}$ data available to perform spectroscopy studies.

\section{Charm-strange spectroscopy}

Several new analyses have been presented at this conference, combining Belle and Belle II data sets. In particular, we presented the results of the observation of $cs$-meson pair production in $\Upsilon(2S)$,  and $cs$-spectroscopy in the continuum. 

\section{Charm-strange from $\Upsilon(2S)$ decays}

The charm--strange sector continues to provide crucial insights into the dynamics of QCD in the heavy–light regime. Several new analyses combining Belle and Belle~II data sets have been presented at this conference. In particular, the observation of $c\bar{s}$ meson pair production in $\Upsilon(2S)$ decays at Belle~\cite{ref_cspair}  represents an important step forward. The analysis, performed on 24.7\,fb$^{-1}$ of data collected at the $\Upsilon(2S)$ resonance, provides new information on the strong interaction dynamics governing heavy–quark bound states and tests theoretical predictions in the charm–strange sector.

Born cross-section measurements have been shown at this conference, showing that strong decays dominate in the $\Upsilon(2S) \rightarrow D_s^{(*)+} D_s^-$ process.

\subsection{The $D_{s0}^*(2317)^+$ puzzle}

The $D_{s0}^*(2317)^+$ remains one of the most intriguing states in the charm–strange spectrum~\cite{ref_Ds2317_BaBar2003, ref_Ds2317_CLEO2003, ref_Ds2317_Belle2004,ref_GodfreyIsgur,  ref_BarnesCloseLipkin, ref_Tetraquark2003, ref_Ds2317_PTEP2024, ref_Ds2317_arXiv2310,  ref_Ds2317_PRD2023}. Its mass is significantly lower than predicted by quark–model calculations, suggesting a possible non-conventional structure, such as a $DK$ molecular component or a mixed $c\bar{s}$–molecule configuration.

A new analysis combining Belle (983\,fb$^{-1}$) and Belle~II (427.9\,fb$^{-1}$) data reconstructs the decay $D_{s0}^*(2317)^+ \to D_s^* \gamma$, with the goal of providing a precise mass measurement and branching fraction determination. 
The combined luminosity of 1.41\,ab$^{-1}$ makes this the most sensitive study to date, offering stringent constraints on theoretical models. The relative BR evaluated quotes the ratio R = BR( $D_{s0}^*(2317)^+ \to D_s^{*+} \gamma$)/BR( $D_{s0}^*(2317)^+ \to D_s^{+} \pi^0 )$ = (7.14 $\pm$ 0.70 $\pm$ 0.23)$\%$. In conventional quark‑model interpretations, the radiative decay is expected to dominate over the isospin‑violating transition, with predicted branching fractions typically in the 10–50$\%$ range, while in the tetraquark or in th DK-molecular model is expected to be 1-10$\%$. This measurement represents a direct test of the internal structure of the   $D_{s0}^*(2317)^+$.

\subsection{From $c\bar{s}$ mesons to tetraquarks: search for $cc\bar{s}\bar{s}$}

The transition from conventional $c\bar{s}$ spectroscopy to the search for multiquark states is a natural extension of the Belle~II spectroscopy physics program. 
Belle has performed the first measurement of the processes $e^+e^- \to D_s^+ D_{s0}^*(2317)^- X$ and $e^+e^- \to D_s^+ D_{s1}(2460)^- X$ in the continuum. No evidence for resonant structures was found up to 980\,fb$^{-1}$. These results provide essential input for refining models of tetraquark production and constrain the possible existence of $cc\bar{s}\bar{s}$ states.

\begin{figure}[t]
  \centering
  \includegraphics[width=0.45\linewidth]{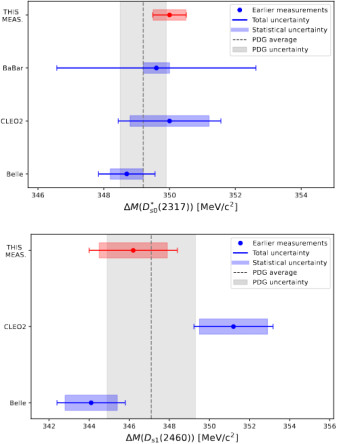}
  \caption{Comparison between the preliminary $\Delta M$ mesurement proposed in this work (red rectangle) and what exists in literature (blue rectangles, from other experiments as indicated). The analysis is presented in Sec. 3.2. }
  \label{fig:Fig1}
\end{figure}

The precise measurement of the mass difference $\Delta M = M(D_{s1}(2460)) - M(D_{s0}^*(2317))$ also offers a sensitive probe of chiral symmetry breaking in heavy–light systems. In this preliminary work we summarize our results in Fig.~\ref{fig:Fig1}, in comparison with literature.

\section{Search for pentaquarks}

\subsection{Search in the $pJ/\psi$ final state}

Belle has performed an inclusive search for pentaquark candidates in the $pJ/\psi$ final state 
using $\Upsilon(1S)$ and $\Upsilon(2S)$ data, corresponding to a total integrated luminosity of 30.5\,fb$^{-1}$, 
supplemented by 89\,fb$^{-1}$ collected 60\,MeV below the $\Upsilon(4S)$ peak~\cite{ref_pcjpsi}. 
The analysis reconstructs $P_c$ candidates inclusively on the recoil of the rest of the event, while suppressing Bhabha background and reflection peaks by imposing $M^2(pJ/\psi) > 10$\,GeV$^2/c^4$.

\begin{figure}[t]
  \centering
  \includegraphics[width=0.45\linewidth]{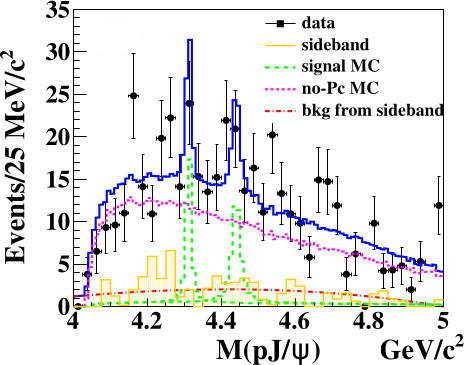}
  \caption{Fit result including $P_c^+ (4312)$,  $P_c^+ (4440)$, and  $P_c^+ (4457)$.  $\Upsilon(1S)$ and $\Upsilon(2S)$ data are used. This anaylsis is introduced in Sec. 4.1. }
  \label{fig:Fig2}
\end{figure}

No significant signal was observed for the $P_c(4312)^+$, $P_c(4440)^+$, or $P_c(4457)^+$ states. These results provide important constraints on the production of hidden-charm pentaquarks in $e^+e^-$ collisions (see Fig.~\ref{fig:Fig2}).

\subsection{Search in the $\Lambda J/\psi$ final state}

\begin{figure}[t]
  \centering
  \includegraphics[width=0.45\linewidth]{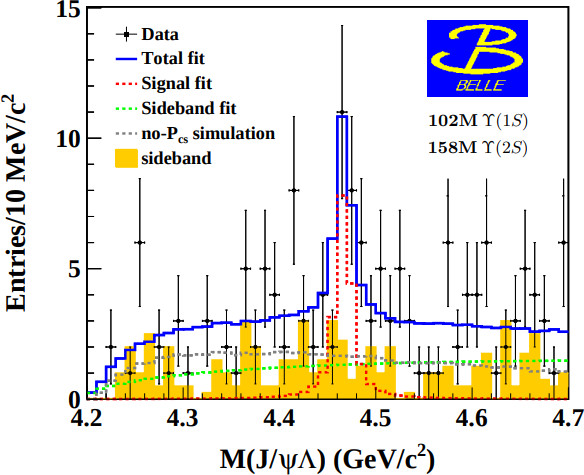}
  \caption{Study of the $J/\psi \Lambda$ invariant mass presented in Sec. 4.2. First evidence for the $P_{cs}(4459)^0$ is shown using $\Upsilon(1S)$ and  $\Upsilon(2S)$ data sets at Belle.}
  \label{fig:Fig4}
\end{figure}

A complementary analysis investigates the $\Lambda J/\psi$ final state in $\Upsilon(1S)$ and $\Upsilon(2S)$ decays~\cite{ref_pcs}. 
This study reports the first observation of the decays $\Upsilon(1S) \to \Lambda J/\psi$ and $\Upsilon(2S) \to \Lambda J/\psi$, 
and finds evidence at the 3.8$\sigma$ level for the $P_{cs}(4459)^0$ state. 
This constitutes the first indication of a strange pentaquark candidate in an $e^+e^-$ environment, providing an important cross-check of the LHCb observations (see Fig.~\ref{fig:Fig4}). The fitted parameters in this analysis are: M = (4471.7 $\pm$ 4.8 $\pm$ 0.6) MeV/$c^2$ and $\Gamma$ = (21.9 $\pm$ 13.1 $\pm$ 2.7) MeV. The excess of events is seen with 3.8$\sigma$ statistical significance. This is the evidence at Belle of the $P_{cs}(4459)^0$, and it is compatible with the former LHCb report on the same resonance.

\section{Baryon spectroscopy}

\subsection{The $\Omega(2012)$}

The $\Omega(2012)$ baryon was first observed by Belle in two-body decays ($\Xi K$)~\cite{ref_omega2012}. A new analysis of three-body decays $\Xi^- \pi^+ K^-$~\cite{ref_omega2012_3body} strongly suggests a quantum number assignment of $J^P = 3/2^-$, having measured the ratio of couplings $\frac{g_3}{g_2}$ = $22.9^{+17.9}_{.22.4} \pm 2.2$. The two-body decay modes have also been remeasured, improving the precision on the branching fractions.

\subsection{Charmed baryons}

\begin{figure}[t]
  \centering
  \includegraphics[width=0.75\linewidth]{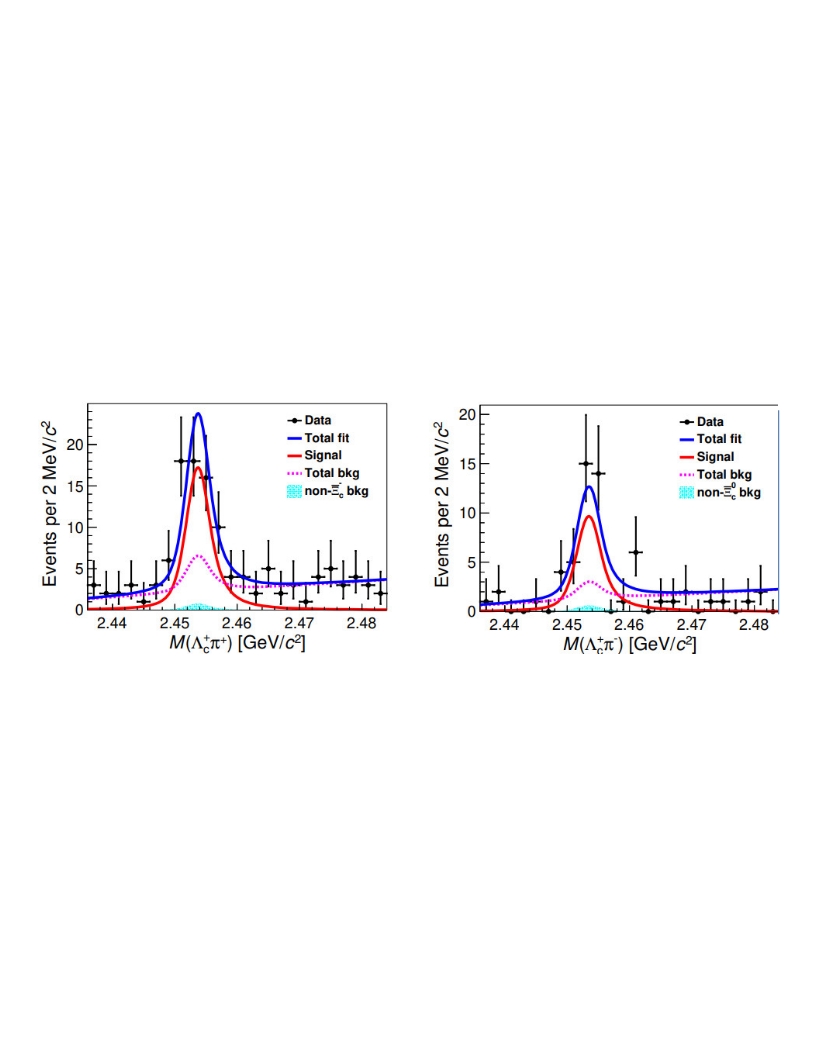}
  \caption{Distributions of $M(\Lambda_c^+ \pi^\pm)$ for 
    $B^+ \rightarrow \Sigma_c(2455)^{++}\,\bar{\Xi}_c^{-}$ (left) and 
    $B^0 \rightarrow \Sigma_c(2455)^{0}\,\bar{\Xi}_c^{0}$ (right).}
  \label{fig:Fig5}
\end{figure}

Belle~II has searched for charmed baryons decaying to $\Lambda_c \eta$ using 364\,fb$^{-1}$ of data, 
finding no significant excess, in agreement with expectations~\cite{ref_lceta}. 
Belle has also studied the substructure of $\Lambda_c^+ \to \Lambda \pi^+ \pi^- \pi^+$ using 980\,fb$^{-1}$, 
observing a clear enhancement near the $NK$ threshold. 
Further data from Belle~II will be essential to determine whether this structure corresponds to a cusp or a genuine resonance. 
A combined Belle + Belle~II analysis has led to the first observation of the decays 
$B^{+/0} \to \Sigma_c(2455)^{++/0}\,\bar{\Xi}_c^{-/0}$, 
with branching fractions consistent with diquark-model expectations (see Fig.~\ref{fig:Fig5}).

\section{Bottomonium physics}

\subsection{Energy scan studies}

Belle~II performed its first energy scan in November 2021, collecting 20\,fb$^{-1}$ of data~\cite{ref_scan2023}. 
The goal was to study the $\Upsilon(10753)$ region and the $B^*B^*$ threshold. 
The scan revealed that the $\Upsilon(10753)$ and $\Upsilon(5S)$ exhibit different line-shape patterns, suggesting a more complex structure than previously assumed. Studies of exclusive processes such as $e^+e^- \to \omega \chi_{bJ}$, and $e^+e^- \to BB$, $BB^*$, $B^*B^*$ have been reported in Refs.~\cite{ref_bbbar1,ref_bbbar2}.

\subsection{Radiative and hadronic transitions}

\begin{figure}[t]
  \centering
  \includegraphics[width=0.55\linewidth]{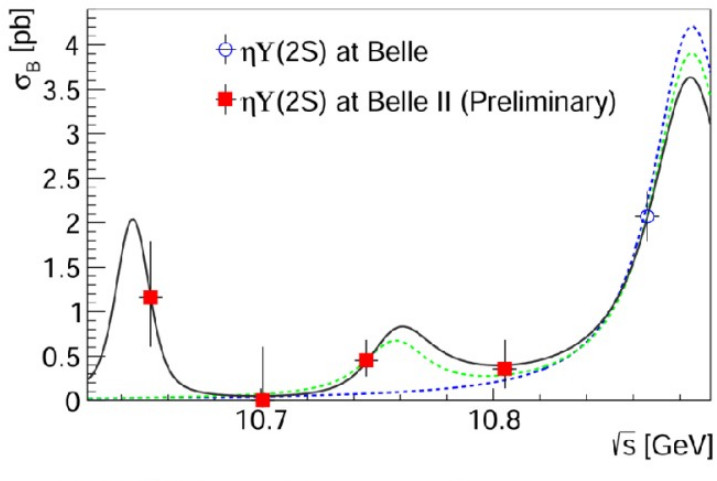}
  \caption{Measured Born cross sections of the $e^+ e^- \rightarrow  \eta \Upsilon(2S)$ process }
  \label{fig:Fig6}
\end{figure}

Belle~II has searched for radiative transitions $e^+e^- \to \gamma \chi_{bJ}$ 
without finding significant signals, indicating that the D-wave component of the $\Upsilon(10753)$ may be small. A recent study 
reports an observation above 6$\sigma$ and suggests that the signal cannot originate solely from the $\Upsilon(10753)$ or $\Upsilon(5S)$. The data indicate the possible presence of a new state near 10.60\,GeV, consistent with earlier evidence in the $B^*B^*$ channel. In this analysis, 4 samples were collected near the $\Upsilon(10753)$. We obtained confirmation of the evidence already found in the $e^+ e^- \rightarrow B^+ B^. $ inelastic decay channel (see Fig.~\ref{fig:Fig6}). 

\section{Precision measurements}

\subsection{Mass difference between $B^0$ and $B^+$}

A new measurement of the mass difference 
$\Delta M = M(B^0) - M(B^+)$ 
combines 571\,fb$^{-1}$ of Belle data with 365\,fb$^{-1}$ from Belle~II~\cite{ref_bmass}. 
The result,

\[
\Delta M = (0.495 \pm 0.024 \pm 0.005)\,\text{MeV},
\]

provides a precise test of isospin symmetry breaking and contributes to improved modeling of $B$-meson decays. 
The analysis also measures the ratio 
$R = \sigma(B^0\bar{B}^0)/\sigma(B^+B^-)$, 
demonstrating that the phase-space assumption used by BaBar is excluded at the 10$\sigma$ level.

\subsection{Contribution to $(g-2)_\mu$}

Belle has recently measured the cross section of $e^+e^- \to \pi^+\pi^-\pi^0$~\cite{ref_gminus2}, providing new input to the hadronic vacuum polarization (HVP) contribution to the muon anomalous magnetic moment. 
This measurement helps reduce the theoretical uncertainty in the Standard Model prediction of $a_\mu$. Using solely the Belle result, we obtained in this analysis $a_\mu^{LO, HPV, 3\pi} (0.62 - 1.8 GeV) = (48.91 \pm 0.25_{stat} \pm 1.07_{sys} ) \times 10^{-10}$, which is 6.5$\%$ higher than the global fit result, with 2.5$\sigma$ statistical significance. This discrepancy corresponds to 10$\%$ of the difference $\Delta a_\mu = a_\mu(Exp) - a_\mu(SM) = 2.5 \times 10^{-10}$, SM = Standard Model theoretical value.

\section{Summary}

Although Belle ceased operations in 2010, its data continue to produce high-impact results. 
The combination of Belle and Belle~II data sets, now exceeding 1.5 ab$^{-1}$, 
has enabled new precision measurements and searches for exotic hadrons. 
Belle~II has already collected 575\,fb$^{-1}$ and will resume data taking in late 2025, 
with a long-term goal of 50\,ab$^{-1}$. 
The results presented here demonstrate the strong potential of the Belle~II program in spectroscopy, 
from multiquark searches to bottomonium physics and precision measurements.

\end{document}